\documentclass[aps,preprint]{revtex4}%
\usepackage{amsfonts}
\usepackage{amsmath}
\usepackage{amssymb}
\usepackage{graphicx}%
\newtheorem{theorem}{Theorem}

\newenvironment{proof}[1][Proof]{\noindent\textbf{#1.} }{\ \rule{0.5em}{0.5em}}
\begin{document}
\preprint{ }
\title{Generating Static Spherically Symmetric Black-holes in Lovelock Gravity}
\author{S. Habib Mazharimousavi}
\email{habib.mazhari@emu.edu.tr}
\author{O. Gurtug}
\email{ozay.gurtug@emu.edu.tr}
\author{M. Halilsoy}
\email{mustafa.halilsoy@emu.edu.tr}
\affiliation{Department of Physics, Eastern Mediterranean University, G. Magusa, North
Cyprus, Mersin 10 - Turkey.}
\pacs{PACS number}

\begin{abstract}
Generalization of a known theorem \ to generate static, spherically symmetric
black-hole solutions in higher dimensional Lovelock gravity is presented.
Particular limits, such as Gauss-Bonnet (GB) and/or Einstein-Hilbert (EH) in
any dimension $N$ yield all the solutions known to date with an
energy-momentum. In our generalization, with special emphasis on the third
order Lovelock gravity, we have found two different class of solutions
characterized by the matter field parameter. Several particular cases are
studied and properties related to asymptotic behaviours are discussed. Our
general solution which covers topological black holes as well, splits
naturally into distinct classes such as Chern-Simon (CS) and Born-Infeld (BI)
in higher dimensions. The occurence of naked singularities are studied and it
is found that, the spacetime behaves nonsingular in quantum mechanical sense
when it is probed with quantum test particles. The theorem is extended to
cover Bertotti-Robinson (BR) type solutions in the presence of the GB
parameter alone. Finally we prove also that extension of the theorem for a
scalar-tensor source of higher dimensions $(N>4)$ fails to work.

\end{abstract}
\maketitle

\section{Introduction}

One of the most interesting features of the string theory is to provide an
arena for higher dimensional space-times. String theory together with higher
dimensions supports also the description of objects known as branes. There is
no doubt that, the most intriguing solution in higher dimensional space-times
is the one that is associated with black holes. The pioneering work in this
regard belong to Boulware and Deser \cite{1}. They obtained the most general
static black hole solutions in Einstein - Gauss - Bonnet (EGB) theory. Recent
studies show that there is a growing interest to find black-hole solutions in
higher dimensional gravity. This task is accomplished by the use of the most
general action that describes black-hole solutions in Einstein-Lovelock
theory\cite{2}. This is the most general theory that hosts higher order
invariants in particular combinations so that field equations remain second
order and therefore do not contain ghosts. Physical properties of the Einstein
- Lovelock theory that admits black holes is analyzed in detail in the
Ref.\cite{3}. Black hole solutions in Lovelock theory is important in the
sense that the higher order curvature terms contribute to the inner structure
of black holes. For example, in 4- dimensional general relativity, inner
(Cauchy) and outer (event) horizons arise in the Reissner-Nordstr\"{o}m black
hole in which the corresponding space-time admits two Killing vectors
orthogonal to each other. However, it is demonstrated in \cite{4} that, double
horizons may arise in the Einstein - Lovelock theory in the absence of the
matter fields as well. Another remarkable aspect of the Lovelock theory is to
provide topological black holes in which the curvature scalar is not
positive\cite{5,6,7}.

In general theory of relativity, static, spherically symmetric space-times
constitute one of the simplest and tractable model as far as the analytic,
exact solutions are concerned. Recently, Salgado \cite{8}, has developed a
theorem, to generate exact black-hole solutions by imposing some conditions on
the energy-momentum tensor. This theorem has been extended to the higher
dimensional spacetimes \cite{9} in which Schwarzschild, Reissner-Nordstr\"{o}m
and global monopole solutions in higher dimensions are particular cases. Later
on, another version of the theorem has been introduced to generate exact
radiative (dynamic) black-hole solutions \cite{10}. This can be interpreted as
the dynamic extension of the Salgado's theorem in which the generalized,
spherically symmetric Type II fluid solution is obtained. Many known solutions
are shown to be the particular limits of this generalized solution. The
extension of this non-static radiative solution in the EGB and Born-Infeld
nonlinear electrodynamic theory is also considered in \cite{11}. More
recently, it has been shown that the $N$-dimensional extension of radiative,
dynamic black-hole solutions are possible as well\cite{12}.

In this paper, we wish to extend Salgado's theorem to the arbitrary
dimensional, Lovelock gravity coupled with matter fields, starting with the
third order. In our case, matter fields couple to the system through an
arbitrary constant parameter. It is shown that, our general solution includes
the well-known solutions in particular limits, namely, the GB and Einstein
limits in any lower dimensional solutions. Besides, in $N=7$, we present a
black hole solution with interesting properties that, depending on the
constant parameters $\alpha_{2}$ (the GB parameter)$,$ $\alpha_{3}$(the third
order Lovelock parameter) and $C$ (the energy momentum parameter), one or two
horizons may develop. Moreover, we emphasize that depending on these
parameters our solutions are either flat or de Sitter / anti - de Sitter
types. We discuss also the behavior of the naked singularity when it is probed
with quantum particles. It is found quantum mechanically that, the classical
timelike curvature singularity at the origin remains nonsingular .

It is a known fact that all standard black hole solutions in spherically
symmetric spacetimes possess a central singularity at $r=0$. Our analysis has
shown for $N=7$ that the third order Lovelock term is effective in removing
the black hole property and leaving the central singularity at $r=0$ as a
\textit{naked} singularity. It can be shown that for $N>7$, higher order
parameters $\alpha_{s}$, $\left(  s>3\right)  $ plays a similar role.
Following the analysis of Ref.\cite{13}\ we can scan the family of black holes
in general Lovelock theory which can be labelled as Chern-Simon (CS),
Born-Infeld (BI) types and those that fit neither scheme. This is related with
the odd/even dimensionality and fine-tuning of the coupling parameter which
plays crucial roles in the thermodynamical properties. We consider the
implication of the Salgado's theorem within the context of Bertotti-Robinson
(BR) type spacetimes. As a final application of the Salgado's theorem we
investigate whether it is applicable for a scalar field in higher dimensions:
It turns out that the theorem is valid only in $N=4$.

The paper is organized as follows: In Sec. $II$ we give a brief summary of the
$N$-dimensional Einstein - third order Lovelock gravity. We present the static
spherically symmetric $7$ - dimensional solution in third order Lovelock
theory \ with physical properties in Sec. $III$. \ In section $IV$ we present
the general solution to the Lovelock gravity that confirms Salgado's theorem .
In Sec. $V,$ we introduce the general form of the theorem concerning BR type
solutions with GB term in any dimension and investigate applicability of the
theorem to the scalar - tensor theory. The paper ends with a conclusion in
Sec. $VI$.

\section{$N$-dimensional Third Order Einstein - Lovelock Gravity.}

The action describing $N$-dimensional third order Einstein - Lovelock gravity
coupled with matter fields is given by;%

\begin{equation}
S=\int dx^{N}\sqrt{-g}\left\{
%TCIMACRO{\tciLaplace}%
%BeginExpansion
\mathcal{L}%
%EndExpansion
_{EH}+\alpha_{2}%
%TCIMACRO{\tciLaplace}%
%BeginExpansion
\mathcal{L}%
%EndExpansion
_{GB}+\alpha_{3}%
%TCIMACRO{\tciLaplace}%
%BeginExpansion
\mathcal{L}%
%EndExpansion
_{L}\right\}  +S_{matter},
\end{equation}
where $%
%TCIMACRO{\tciLaplace}%
%BeginExpansion
\mathcal{L}%
%EndExpansion
_{EH}$ (the first order, or Einstein-Hilbert term), $%
%TCIMACRO{\tciLaplace}%
%BeginExpansion
\mathcal{L}%
%EndExpansion
_{GB}$ (the second order, or Gauss-Bonnet term) and $%
%TCIMACRO{\tciLaplace}%
%BeginExpansion
\mathcal{L}%
%EndExpansion
_{L}$ ( the third order Lovelock term) are defined as follows
\begin{align}%
%TCIMACRO{\tciLaplace}%
%BeginExpansion
\mathcal{L}%
%EndExpansion
_{EH}  &  =R,\nonumber\\%
%TCIMACRO{\tciLaplace}%
%BeginExpansion
\mathcal{L}%
%EndExpansion
_{GB}  &  =R_{\mu\nu\gamma\delta}R^{\mu\nu\gamma\delta}-4R_{\mu\nu}R^{\mu\nu
}+R^{2},\\%
%TCIMACRO{\tciLaplace}%
%BeginExpansion
\mathcal{L}%
%EndExpansion
_{L}  &  =2R^{\mu\nu\sigma\kappa}R_{\sigma\kappa\rho\tau}R_{\text{ \ \ }\mu
\nu}^{\rho\tau}+8R_{\text{ \ \ }\sigma\rho}^{\mu\nu}R_{\text{ \ \ }\nu\tau
}^{\sigma\kappa}R_{\text{ \ \ }\mu\kappa}^{\rho\tau}+24R^{\mu\nu\sigma\kappa
}R_{\sigma\kappa\nu\rho}R_{\text{ \ }\mu}^{\rho}\nonumber\\
&  +3RR^{\mu\nu\sigma\kappa}R_{\sigma\kappa\mu\nu}+24R^{\mu\nu\sigma\kappa
}R_{\sigma\mu}R_{\kappa\nu}+16R^{\mu\nu}R_{\nu\sigma}R_{\text{ \ }\mu}%
^{\sigma}-12RR^{\mu\nu}R_{\mu\nu}+R^{3}.\nonumber
\end{align}
The constants $\alpha_{2}$ and $\alpha_{3}$ stand for arbitrary constants
whereas $S_{matter}$ represents the action of the matter fields. We recall
that for $S_{matter}=0,$ $%
%TCIMACRO{\tciLaplace}%
%BeginExpansion
\mathcal{L}%
%EndExpansion
_{GB}$ and $%
%TCIMACRO{\tciLaplace}%
%BeginExpansion
\mathcal{L}%
%EndExpansion
_{L}$ terms become meaningful only for $N\geq5$ and $N\geq7,$ respectively.
Variation of the action with respect to the metric tensor $g_{\mu\nu}$ yields
the field equations in the form%

\begin{equation}
G_{\mu\nu}^{EH}+\alpha_{2}G_{\mu\nu}^{GB}+\alpha_{3}G_{\mu\nu}^{L}=T_{\mu\nu},
\end{equation}
where $T_{\mu\nu}$ is the energy-momentum tensor representing the matter
fields. $G_{\mu\nu}^{EH}$ is the Einstein tensor, and $G_{\mu\nu}^{GB}$ and
$G_{\mu\nu}^{L}$ are given as:%

\begin{equation}
G_{\mu\nu}^{GB}=2(-R_{\mu\sigma\kappa\tau}R_{\text{ \ \ \ \ }\nu}^{\kappa
\tau\sigma}-2R_{\mu\rho\nu\sigma}R^{\rho\sigma}-2R_{\mu\sigma}R_{\text{ \ }%
\nu}^{\sigma}+RR_{\mu\nu})-\frac{1}{2}%
%TCIMACRO{\tciLaplace}%
%BeginExpansion
\mathcal{L}%
%EndExpansion
_{GB}g_{\mu\nu},
\end{equation}

\begin{align}
G_{\mu\nu}^{L}  &  =-3(4R^{\tau\rho\sigma\kappa}R_{\sigma\kappa\lambda\rho
}R_{\text{ \ \ }\nu\tau\mu}^{\lambda}-8R_{\text{ \ \ }\lambda\sigma}^{\tau
\rho}R_{\text{ \ \ }\tau\mu}^{\sigma\kappa}R_{\text{ \ \ }\nu\rho\kappa
}^{\lambda}+2R_{\text{ }\nu}^{\text{ \ \ }\tau\sigma\kappa}R_{\sigma
\kappa\lambda\rho}R_{\text{ \ \ }\tau\mu}^{\lambda\rho}\\
&  -R^{\tau\rho\sigma\kappa}R_{\sigma\kappa\lambda\rho}R_{\nu\mu}+8R_{\text{
\ \ }\nu\sigma\rho}^{\tau}R_{\text{ \ \ }\tau\nu}^{\sigma\kappa}R_{\text{
\ \ }\kappa}^{\rho}+8R_{\text{ \ \ }\nu\tau\kappa}^{\sigma}R_{\text{
\ \ }\sigma\mu}^{\tau\rho}R_{\text{ \ \ }\rho}^{\kappa}\nonumber\\
&  +4R_{\nu}^{\text{ \ }\tau\sigma\kappa}R_{\sigma\kappa\mu\rho}R_{\text{
\ \ }\tau}^{\rho}-4R_{\nu}^{\text{ \ }\tau\sigma\kappa}R_{\sigma\kappa\tau
\rho}R_{\text{ \ \ }\mu}^{\rho}+4R^{\tau\rho\sigma\kappa}R_{\sigma\kappa
\tau\mu}R_{\nu\rho}+2RR_{\nu}^{\text{ \ }\kappa\tau\rho}R_{\tau\rho\kappa\mu
}\nonumber\\
&  +8R_{\text{ \ \ }\nu\mu\rho}^{\tau}R_{\text{ \ \ }\sigma}^{\rho}R_{\text{
\ \ }\tau}^{\sigma}-8R_{\text{ \ \ }\nu\tau\rho}^{\sigma}R_{\text{ \ \ }%
\sigma}^{\tau}R_{\text{ \ \ }\mu}^{\rho}-8R_{\text{ \ \ }\sigma\mu}^{\tau\rho
}R_{\text{ \ \ }\tau}^{\sigma}R_{\text{ }\nu\rho}-4RR_{\text{ \ \ }\nu\mu\rho
}^{\tau}R_{\text{ \ \ }\tau}^{\rho}\nonumber\\
&  +4R^{\tau\rho}R_{\rho\tau}R_{\nu\mu}-8R_{\text{ \ \ }\mu}^{\tau}R_{\text{
}\tau\rho}R_{\text{ \ \ }\mu}^{\rho}+4RR_{\text{ }\nu\rho}R_{\text{ \ \ }\mu
}^{\rho}-R^{2}R_{\nu\mu})-\frac{1}{2}%
%TCIMACRO{\tciLaplace}%
%BeginExpansion
\mathcal{L}%
%EndExpansion
_{L}g_{\mu\nu}.\nonumber
\end{align}

\section{The Static Solution.}

Generalization of the Salgado's theorem to the third order Lovelock theory
together with arbitrary matter fields is as follows:

\begin{theorem}
Let ($M,g_{ab})$ be a N-dimensional space-time with sign $(g_{ab})=N-2,$
$N\geq3$, such that : (1) it is static and spherically symmetric, (2) it
satisfies the Einstein field equations, (3) the energy momentum tensor
$T^{ab}$ satisfies the conditions $T_{r}^{r}=T_{t}^{t}$ and $T_{\theta_{i}%
}^{\theta_{i}}=kT_{r}^{r}$ $(1\leq i\leq n-2,$ $k=$constant $\epsilon$ $%
%TCIMACRO{\U{211d} }%
%BeginExpansion
\mathbb{R}
%EndExpansion
$ $)$, (4 ) it possess a regular Killing horizon or a regular origin. Then,
the metric of the space-time is given by
\end{theorem}

\begin{equation}
ds^{2}=-f(r)dt^{2}+f(r)^{-1}dr^{2}+r^{2}d\sigma_{n}^{2},
\end{equation}
\textit{where}%
\[
d\sigma_{n}^{2}=\left\{
\begin{tabular}
[c]{ll}%
$d\theta_{1}^{2}+\sin^{2}\theta_{1}\sum_{i=2}^{n}%
%TCIMACRO{\dprod \limits_{j=2}^{i-1}}%
%BeginExpansion
{\displaystyle\prod\limits_{j=2}^{i-1}}
%EndExpansion
\sin^{2}\theta_{j}d\theta_{i}^{2},$ & $0\leq\theta_{n}\leq2\pi,0\leq\theta
_{i}\leq\pi,\text{ \ }1\leq i\leq n-1,\text{for }\chi=1,$\\
$\sum_{i=1}^{n}d\theta_{i}^{2},$ & $\text{\ }0\leq\theta_{i}\leq2\pi,\text{
\ \ \ for \ \ }\chi=0,$\\
$d\theta_{1}^{2}+\sinh^{2}\theta_{1}\sum_{i=2}^{n}%
%TCIMACRO{\dprod \limits_{j=2}^{i-1}}%
%BeginExpansion
{\displaystyle\prod\limits_{j=2}^{i-1}}
%EndExpansion
\sin^{2}\theta_{j}d\theta_{i}^{2},$ & $\text{\ }0\leq\theta_{n}\leq2\pi
,0\leq\theta_{i}\leq\pi,\text{ \ }1\leq i\leq n-1,\text{for }\chi=-1,$%
\end{tabular}
\ \right.  \text{ \ \ \ \ }%
\]
\textit{stands for the line element of the }$n-$\textit{dimensional base
manifold \ }$\Sigma$ \textit{which is assumed to be compact, without boundary,
and of constant curvature }$n\left(  n-1\right)  \chi$ \textit{that without
loss of the generality, one may take }$\chi=$\textit{ }$\pm1,0$\textit{. This
implies that the surface is locally isometric to the sphere }$S^{n}$\textit{,
flat space }$R^{n}$\textit{, or to the hyperbolic manifold }$H^{n}$\textit{
for }$\chi=1$\textit{; }$0$\textit{; }$-1$\textit{, respectively.and the
energy momentum tensor in general is in the following form }%
\begin{equation}
T_{\mu\left(  Diag.\right)  }^{\nu}=\frac{C}{r^{n\left(  1-k\right)  }}\left[
1,1,k,...,k\right]  ,
\end{equation}
\textit{in which }$C$\textit{ is an integration constant.}

\begin{proof}
From the hypothesis (1), the related spacetime can be described by the metric,%
\begin{equation}
ds^{2}=-N^{2}(r)dt^{2}+A^{2}(r)dr^{2}+r^{2}d\Omega_{n}^{2}.
\end{equation}
Hypothesis (2), implies that this metric must satisfy the Einstein-Lovelock
equations described by,%
\begin{equation}
\mathcal{G}_{\mu\nu}=G_{\mu\nu}^{EH}+\alpha_{2}G_{\mu\nu}^{GB}+\alpha
_{3}G_{\mu\nu}^{L}=T_{\mu\nu}.
\end{equation}
From the hypothesis (3), $T_{\text{ }t}^{t}-T_{\text{ }r}^{r}=0$ and hence,
one finds $\mathcal{G}_{\text{ }t}^{t}-\mathcal{G}_{\text{ }r}^{r}=0$ whose
explicit form on integration gives $\mid g_{00}g_{11}\mid=C_{0}=$ constant,
and it remains to choose the time scale at infinity to make this constant
equal to unity. This leads to choose the metric functions such that,%
\begin{equation}
N^{2}(r)=f(r)\text{ \ \ \ \ \ \ \ \ \ and \ \ \ \ \ \ \ \ \ \ }A^{2}%
(r)=f^{-1}(r).
\end{equation}
Among others, the $rr$-component of the Eq. (9) is the simplest one and \ can
be written as,%
\begin{align}
T_{\text{ }r}^{r}  &  =\frac{n}{2r^{6}}\{\left[  r^{5}-2\widetilde{\alpha}%
_{2}r^{3}g(r)+3\widetilde{\alpha}_{3}rg(r)^{2}\right]  g^{\prime}\left(
r\right)  +\left(  n-1\right)  r^{4}g\left(  r\right) \\
&  -\left(  n-3\right)  \widetilde{\alpha}_{2}r^{2}g(r)^{2}+\left(
n-5\right)  \widetilde{\alpha}_{3}g(r)^{3}\}\nonumber
\end{align}
in which a prime denotes derivative\ with respect to $r$, $g(r)=f(r)-\chi$,
$\widetilde{\alpha}_{2}=\left(  n-1\right)  \left(  n-2\right)  \alpha_{2}$,
$\widetilde{\alpha}_{3}=\left(  n-1\right)  \left(  n-2\right)  \left(
n-3\right)  \left(  n-4\right)  \alpha_{3}$ and $n=N-2.$ From the conservation
laws $\nabla_{\mu}T_{\text{ \ }\nu}^{\mu}=0,$ \ we have,%
\begin{equation}
\partial_{r}T_{\text{ }r}^{r}=\frac{1}{2f}\left(  T_{\text{ }t}^{t}-T_{\text{
}r}^{r}\right)  \frac{\partial f}{\partial r}-\frac{n}{r}\left(  T_{\text{ }%
r}^{r}-T_{\text{ }\theta_{1}}^{\theta_{1}}\right)  .
\end{equation}
Using hypothesis (3), this equation reduces to,%
\begin{equation}
\partial_{r}T_{\text{ }r}^{r}=-\frac{n}{r}\left(  T_{\text{ }r}^{r}-T_{\text{
}\theta_{1}}^{\theta_{1}}\right)  ,
\end{equation}
whose integration gives,%
\begin{equation}
T_{\text{ }r}^{r}=\frac{C}{r^{n\left(  1-k\right)  }}.
\end{equation}
These results can be combined, so that we have the diagonal $T_{\text{ }\nu
}^{\mu}$ as,%
\begin{equation}
T_{\text{ }\nu\left(  Diag.\right)  }^{\mu}=\frac{C}{r^{n\left(  1-k\right)
}}\left[  1,1,k,...,k\right]  .
\end{equation}
The general solution of the Eq.(9) under (11) in any dimension is obtained for
two different broad classes $A$ and $B$ as follows.
\end{proof}

\textbf{Class }$A$\textbf{: }The solution in this class is categorized
according to the energy-momentum parameter \ $k\neq-\frac{1}{n}$, given by;%

\begin{equation}
f(r)=\chi+\frac{\widetilde{\alpha}_{2}}{3\widetilde{\alpha}_{3}}%
r^{2}-2G\left(  1-\frac{\widetilde{\alpha}_{2}^{2}}{3\widetilde{\alpha}_{3}%
}\right)  r^{n-1}\xi_{2}^{-1/3}+\frac{1}{6\widetilde{\alpha}_{3}Gr^{n-5}}%
\xi_{2}^{1/3},\text{ \ \ \ \ for \ \ \ \ \ }k\neq-\frac{1}{n},\text{\ \ \ \ }%
\end{equation}
in which,%

\begin{align}
\xi_{2}  &  =r^{n-9}\left[  216C\widetilde{\alpha}_{3}^{2}r^{n\left(
1+k\right)  }+8G\widetilde{\alpha}_{2}\left(  \widetilde{\alpha}_{2}^{2}%
-\frac{9}{2}\widetilde{\alpha}_{3}\right)  r^{2n}\right.  +\\
&  \left.  12\widetilde{\alpha}_{3}r^{4}\left(  \sqrt{3\xi_{1}}-9\frac
{mG\widetilde{\alpha}_{3}}{n}r^{n-5}\right)  \right]  G^{2},\nonumber\\
\xi_{1}  &  =108C^{2}\widetilde{\alpha}_{3}^{2}r^{2n\left(  1+k\right)
-8}+\frac{G}{n}\left\{  r^{n\left(  3+k\right)  -8}\left(  8n\widetilde
{\alpha}_{2}^{2}C\left(  \widetilde{\alpha}_{2}^{2}-\frac{9}{2}\widetilde
{\alpha}_{3}\right)  +108mC\widetilde{\alpha}_{3}^{2}r^{-n-1}\right)  \right.
-\nonumber\\
&  \left.  \frac{G}{n}r^{2\left(  n-2\right)  }\left[  4mn\widetilde{\alpha
}_{2}r^{n-2}\left(  \widetilde{\alpha}_{2}^{2}-\frac{9}{2}\widetilde{\alpha
}_{3}\right)  +n^{2}\left(  \widetilde{\alpha}_{2}^{2}-4\widetilde{\alpha}%
_{3}\right)  r^{2\left(  n-2\right)  }-27\widetilde{\alpha}_{3}^{2}m^{2}%
r^{-6}\right]  \right\}  ,\nonumber\\
G  &  =n\left(  nk+1\right)  .\nonumber
\end{align}
\textbf{Class }$B$\textbf{: \ }This class represents the solution that belongs
to the energy-momentum parameter $k=-\frac{1}{n}$ and is given by;
\begin{equation}
f(r)=\chi+\frac{\widetilde{\alpha}_{2}}{3\widetilde{\alpha}_{3}}%
r^{2}-2n\left(  1-\frac{\widetilde{\alpha}_{2}^{2}}{3\widetilde{\alpha}_{3}%
}\right)  r^{n+5}\zeta_{2}^{-1/3}+\frac{1}{6\widetilde{\alpha}_{3}nr^{n+1}%
}\zeta_{2}^{1/3},\text{ \ \ \ \ \ \ \ for \ \ \ \ \ \ \ }k=-\frac{1}{n},
\end{equation}
where%

\begin{align}
\zeta_{2} &  =n^{2}r^{2\left(  n+3\right)  }\left[  216\widetilde{\alpha}%
_{3}^{2}r^{2}\left(  C\ln r-\frac{m}{2}\right)  +8\widetilde{\alpha}%
_{2}\left(  \widetilde{\alpha}_{2}^{2}-\frac{9}{2}\widetilde{\alpha}%
_{3}\right)  r^{n+3}+12\widetilde{\alpha}_{3}\sqrt{3\zeta_{1}}\right]  ,\\
\zeta_{1} &  =-n^{2}\left(  \widetilde{\alpha}_{2}^{2}-4\widetilde{\alpha}%
_{3}\right)  r^{2\left(  n+3\right)  }+27\left(  m-2C\ln r\right)
\times\nonumber\\
&  \left[  -\frac{4}{27}n\widetilde{\alpha}_{2}r^{n+5}\left(  \widetilde
{\alpha}_{2}^{2}-\frac{9}{2}\widetilde{\alpha}_{3}\right)  +r^{4}%
\widetilde{\alpha}_{3}^{2}\left(  m-2C\ln r\right)  \right]  .\nonumber
\end{align}
In both classes $A$ and $B$, $m$ is an integration constant that may be
related with the mass. We note that in (n+2)-dimensions for $k=-1$ and
$C=-q^{2},$ class $A$ overlaps with the solution in \cite{4} . Also by setting
$k=\frac{n-4}{n}$ and $C=-\frac{1}{2}n\left(  n-1\right)  Q^{2}$ one recovers
the Einstein-Yang-Mills (EYM) black hole solutions \cite{14,15}. Another case
also in class $A$ is to take $k=1$ which leads us to a perfect-fluid type
energy momentum tensor with constant energy density $(\rho)$, tension $(\tau)$
and pressure $(p)$, such that $C=\rho=\tau=p$. We wish to remark that with the
choices of $m=C=\widetilde{\alpha}_{2}=0,$ leaving behind only $\widetilde
{\alpha}_{3}\neq0,$ leads to the flat space $f(r)=\chi.$ This implies that the
presence of $\widetilde{\alpha}_{3}\neq0$, alone amounts to none other than
the trivial contribution except when $m\neq0$, and/or $C\neq0.$

We observe from the class $B$ which is a new solution that, in any higher
dimension it is possible to have a logarithmic term. We recall that such
solutions were encountered in $5$-dimensional EYM \cite{14} theories. As we
demonstrate in Eqs.(18-19) similar solutions are also possible in
Einstein-Third Order Lovelock gravity. However, their physical interpretation
for higher than $5$- dimensional cases needs further investigation.

\subsection{Properties of the General Solution.}

Since the foregoing solutions are complicated enough for physical
interpretation, we prefer to relate the constants $\widetilde{\alpha}_{2}$ and
$\widetilde{\alpha}_{3}$ in such a way that $3\widetilde{\alpha}%
_{3}=\widetilde{\alpha}_{2}^{2}$ . Let us note that the particular combination
$\widetilde{\alpha}_{2}^{2}-3\widetilde{\alpha}_{3}$ arises naturally in the
formalism. Choosing this to vanish seems to be the easiest simplifying
assumption which accounts for both of the parameters. Given the complexity of
the theory this choice doesn't sacrifice much from the essence of the Lovelock
theory. This choice simplifies the above results to; for class $A$,%

\begin{align}
f(r)  &  =\chi+\frac{1}{2\widetilde{\alpha}_{2}^{2}Gr^{n-5}}\left(  \xi
_{2}^{1/3}+2G\widetilde{\alpha}_{2}r^{n-3}\right)  \text{ \ \ \ \ \ \ \ for
\ \ \ \ }k\neq-\frac{1}{n}\\
\xi_{2}  &  =4\widetilde{\alpha}_{2}^{2}\left[  6r^{n\left(  2+k\right)
-9}C\widetilde{\alpha}_{2}^{2}-3\widetilde{\alpha}_{2}^{2}\frac{G}%
{n}mr^{2\left(  n-5\right)  }-G\widetilde{\alpha}_{2}r^{3\left(  n-3\right)
}+\sqrt{3\xi_{1}}r^{n-5}\right]  G^{2}\nonumber\\
\xi_{1}  &  =3\left\{  4r^{2\left(  n-4+nk\right)  }C^{2}\widetilde{\alpha
}_{2}^{2}+\left[  4\widetilde{\alpha}_{2}Cr^{n\left(  k+2\right)  }\left(
\frac{nr^{n-8}}{3}+r^{-9}\widetilde{\alpha}_{2}m\right)  \right.  \right.
+\nonumber\\
&  \left.  \left.  \left(  \frac{2m}{3}n\widetilde{\alpha}_{2}r^{3n-9}%
+\widetilde{\alpha}_{2}^{2}r^{2n-10}m^{2}+\frac{n^{2}r^{4\left(  n-2\right)
}}{9}\right)  \left(  nk+1\right)  \right]  \left(  nk+1\right)  \right\}
.\nonumber
\end{align}
and class $B$:%

\begin{align}
f(r)  &  =\chi+\frac{r^{2}}{\widetilde{\alpha}_{2}}+\frac{1}{2n\widetilde
{\alpha}_{2}^{2}r^{n+1}}\zeta_{2}^{1/3},\text{ \ \ \ \ \ \ \ for
\ \ \ \ \ \ \ }k=-\frac{1}{n},\\
\zeta_{2}  &  =4n^{2}\widetilde{\alpha}_{2}^{2}r^{2\left(  n+3\right)
}\left[  6\widetilde{\alpha}_{2}^{2}r^{2}\left(  C\ln r-\frac{m}{2}\right)
-n\widetilde{\alpha}_{2}r^{n+3}+\sqrt{3\zeta_{1}}\right]  ,\nonumber\\
\zeta_{1}  &  =\frac{\widetilde{\alpha}_{2}^{2}}{3}\left\{  n^{2}r^{2\left(
n+3\right)  }+9\left(  m-2C\ln r\right)  \widetilde{\alpha}_{2}\left[
\frac{2}{3}nr^{n+5}+r^{4}\widetilde{\alpha}_{2}\left(  m-2C\ln r\right)
\right]  \right\}  ,\nonumber
\end{align}
respectively. The matter fields are coupled to the system through the constant
parameters $C$ and $k.$ The inclusion of the matter fields must satisfy
certain energy conditions as far as the physically acceptable solutions are
concerned. These energy conditions are given in Appendix. According to these
conditions our general solution satisfies the Dominant Energy Condition (DEC)
and restricts the constant parameters as $C\leq0$\ and $\ -1\leq k\leq0$,
while the causality condition bounds the parameter $k$ further to $-1\leq
k\leq-\frac{1}{n}.$

We wish to underline the special case when $k=\frac{1}{n}$. The general
solution for this particular case is, ( from class $A$, either through a
tedious reduction procedure or directly from Eq. (16))%

\begin{equation}
f(r)=\chi+\frac{r^{2}}{\widetilde{\alpha}_{2}}\left\{  1-\sqrt[3]%
{1-\frac{3C\widetilde{\alpha}_{2}}{nr^{n-1}}+\frac{3m\widetilde{\alpha}_{2}%
}{nr^{n+1}}}\right\}  .
\end{equation}
This is a black hole solution with horizon $r_{h}$ which can be found from
$f(r_{h})=0.$ This implies,%

\begin{equation}
nr_{h}^{n-5}\widetilde{\alpha}_{2}^{2}+3nr_{h}^{n-3}\widetilde{\alpha}%
_{2}+3nr_{h}^{n-1}+3Cr_{h}^{2}-3m=0.
\end{equation}
It is important to note that this solution (22) does satisfy only the Weak
Energy Condition (WEC), and therefore the limitations on the constant
parameters are given by $C\leq0$ \ and \ $k\leq1.$ Although this particular
case yields a negative pressure (see Appendix) which may contribute to the
accelerated expansion of the universe, the fact that the Dominant (DEC) and
Strong energy conditions (SEC) and causality are violated limits its applicability.

\subsubsection{Seven - Dimensional Case.}

As it was clarified in Ref.\cite{4}, the physically acceptable description of
the solution obtained in the third order Lovelock gravity is to reduce the
general solution to a seven - dimensional spacetime. The seven - dimensional
spacetimes represent the most general Lagrangian producing second order field
equations, as in the four - dimensional gravity the Einstein - Hilbert action
constitutes the most general such Lagrangian. Further, $N=7$ belongs at the
same time to the class of the CS family of black holes\cite{13}. Hence the
class $A$ solution reduces to;%

\begin{align}
f(r)  &  =\chi+\frac{r^{2}}{\widetilde{\alpha}_{2}}+\frac{\xi_{2}^{1/3}%
}{10\left(  5k+1\right)  \widetilde{\alpha}_{2}^{2}}\text{ \ \ \ \ \ \ \ for
\ \ \ \ \ \ }k\neq-\frac{1}{5}\\
\xi_{2}  &  =100\widetilde{\alpha}_{2}^{2}\left(  5k+1\right)  ^{2}\left[
\sqrt{3\xi_{1}}+6C\widetilde{\alpha}_{2}^{2}r^{5k+1}-\left(  5k+1\right)
\left(  5r^{6}+3m\widetilde{\alpha}_{2}\right)  \right]  ,\nonumber\\
\xi_{1}  &  =-12\widetilde{\alpha}_{2}^{2}\left\{  \left(  5k+1\right)
\left(  m\widetilde{\alpha}_{2}+\frac{5}{3}r^{6}\right)  \left[
C\widetilde{\alpha}_{2}r^{5k+1}-\frac{1}{4}\left(  m\widetilde{\alpha}%
_{2}+\frac{5}{3}r^{6}\right)  \left(  5k+1\right)  \right]  \right.
\nonumber\\
&  \left.  -\widetilde{\alpha}_{2}^{2}C^{2}r^{2\left(  5k+1\right)  }\right\}
,\nonumber
\end{align}

This result generalizes the formerly obtained solutions for any choice of
matter fields upon choosing specific values for $k$ and $C$. For example, if
we choose $k=-1$ and $C=-q^{2},$we recover the solution obtained in \cite{4}.

The other solution ( class $B$ ) becomes%

\begin{align}
f(r)  &  =\chi+\frac{r^{2}}{\widetilde{\alpha}_{2}}+\frac{\zeta_{2}^{1/3}%
}{10\widetilde{\alpha}_{2}^{2}r^{6}},\text{ \ \ \ \ \ \ \ for \ \ \ \ \ \ \ }%
k=-\frac{1}{5},\\
\zeta_{2}  &  =100\widetilde{\alpha}_{2}^{2}r^{16}\left[  6\widetilde{\alpha
}_{2}^{2}r^{2}\left(  C\ln r-\frac{m}{2}\right)  -5\widetilde{\alpha}_{2}%
r^{8}+\sqrt{3\zeta_{1}}\right]  ,\nonumber\\
\zeta_{1}  &  =\widetilde{\alpha}_{2}^{2}r^{4}\{\left[  \widetilde{\alpha}%
_{2}\left(  m-2C\ln r\right)  +5r^{6}\right]  ^{2}-\frac{50}{3}r^{12}%
\}.\nonumber
\end{align}
It can be checked that depending on the signs of $\widetilde{\alpha}_{2}$ and
$\zeta_{2}$ we may have both cases of black hole and non - black hole solutions.

\subsection{ Asymptotic Behavior of the General Solutions For $N=7$:}

For class \textit{A, }the matter field parameter is bounded by $5k+1<6$: The
asymptotic behavior ( as $r\rightarrow\infty$) of class $A$ solution is
investigated for the following possible conditions. According to these
conditions the solutions are given below.

\textit{Case 1}: $\widetilde{\alpha}_{2}\neq0$ $\ $and $\widetilde{\alpha}%
_{3}$ $\neq0.$ The general solution for this case is rather complicated so
that we prefer to give only its asymptotic form,%

\begin{equation}
f(r)\simeq\chi+\Lambda_{eff}r^{2},
\end{equation}
where%

\[
\Lambda_{eff}=\frac{\sqrt[3]{2}\left(  \widetilde{\alpha}_{2}^{2}%
-3\widetilde{\alpha}_{3}\right)  }{3\widetilde{\alpha}_{3}\sqrt[3]{\delta}%
}+\frac{1}{6\widetilde{\alpha}_{3}}\left(  2\widetilde{\alpha}_{2}%
+\sqrt[3]{4\delta}\right)  ,\text{ \ \ \ \ \ \ \ \ }%
\]
in which $\delta=3\widetilde{\alpha}_{3}\sqrt{3}\sqrt{4\widetilde{\alpha}%
_{3}-\widetilde{\alpha}_{2}^{2}}$ $-9\widetilde{\alpha}_{2}\widetilde{\alpha
}_{3}$\ \ $+2$\ $\widetilde{\alpha}_{2}^{3}.$ This is nothing but a de-Sitter
(anti - de Sitter) like behavior.

\textit{Case 2}: $\widetilde{\alpha}_{2}=0$ and $\widetilde{\alpha}_{3}\neq0.$
The general solution now takes the form ( from Eq.s (16) and (17)),%

\begin{equation}
f(r)=\chi-\frac{10\left(  1+5k\right)  }{\sqrt[3]{\xi_{2}}}r^{4}%
+\frac{\sqrt[3]{\xi_{2}}}{30\left(  1+5k\right)  \widetilde{\alpha}_{3}},
\end{equation}
where%

\begin{align*}
\xi_{2}  &  =\widetilde{\alpha}_{3}^{2}\left(  5k+1\right)  ^{2}\left[
5400r^{5k+1}+300\sqrt{3\xi_{1}}-40500\left(  5k+1\right)  m\right]  ,\\
\xi_{1}  &  =-1620\left(  5k+1\right)  mCr^{5k+1}+108C^{2}r^{2\left(
5k+1\right)  }+6075\left(  5k+1\right)  ^{2}\left(  m^{2}+\frac{4}%
{243\widetilde{\alpha}_{3}}r^{12}\right)
\end{align*}
This solution asymptotically behaves as,%

\begin{equation}
f(r)\simeq\chi.
\end{equation}
which implies a flat space. The physical implications of this particular case
will be explored in the next section.

\textit{Case 3}: \ $\widetilde{\alpha}_{2}\neq0$ and $\widetilde{\alpha}%
_{3}=0.$ The general solution is,%

\begin{equation}
f(r)=\chi+\frac{5r^{3}\left(  5k+1\right)  \pm10\sqrt{\frac{\left(
5k+1\right)  }{5}\left[  15\left(  5k+1\right)  \left(  \frac{r^{6}}%
{12}+M\ \widetilde{\alpha}_{2}\right)  -2\ \widetilde{\alpha}_{2}%
Cr^{5k+1}\right]  }}{5\widetilde{\alpha}_{2}\left(  5k+1\right)  r},
\end{equation}
whose asymptotic behavior is%

\begin{equation}
f(r)\simeq\chi+\frac{r^{2}}{\widetilde{\alpha}_{2}}\left(  1\pm1\right)  .
\end{equation}

\textit{Case 4}: $3\widetilde{\alpha}_{3}=\widetilde{\alpha}_{2}^{2}.$ The
general solution for this case is,%

\begin{align}
f(r)  &  =\chi+\frac{r^{2}}{\widetilde{\alpha}_{2}}+\\
&  \frac{1}{5\widetilde{\alpha}_{2}\left(  5k+1\right)  }\sqrt[3]{\left(
5k+1\right)  ^{2}\left[  150\widetilde{\alpha}_{2}Cr^{5k+1}-5625\left(
\widetilde{\alpha}_{2}M+\frac{r^{6}}{9}\right)  \left(  \frac{5k+1}{5}\right)
\right]  },\nonumber
\end{align}
The asymptotic behavior is,%

\begin{equation}
f(r)\simeq\chi.
\end{equation}
i.e. flat.

For class \textit{B}:

\textit{Case 1}: $\widetilde{\alpha}_{2}\neq0$ $\ $and $\widetilde{\alpha}%
_{3}$ $\neq0.$ Due to the complexity of the general solution, we prefer to
give only the asymptotic solution,%

\begin{equation}
f(r)\simeq\chi+\Lambda_{eff}r^{2},
\end{equation}
where%

\[
\Lambda_{eff}=\frac{\widetilde{\alpha}_{2}}{3\widetilde{\alpha}_{3}}+\frac
{1}{6\widetilde{\alpha}_{3}}\sqrt[3]{\delta}+\frac{2\left(  \widetilde{\alpha
}_{2}^{2}-3\widetilde{\alpha}_{3}\right)  }{3\widetilde{\alpha}_{3}%
\sqrt[3]{\delta}}\text{ ,\ \ \ \ }%
\]
in which \ $\delta=8$\ $\widetilde{\alpha}_{2}^{3}-36\widetilde{\alpha}%
_{2}\widetilde{\alpha}_{3}$\ \ $+12\sqrt{3}\widetilde{\alpha}_{3}%
\sqrt{4\widetilde{\alpha}_{3}-\widetilde{\alpha}_{2}^{2}}.$

\textit{Case 2}: $\widetilde{\alpha}_{2}=0$ and $\widetilde{\alpha}_{3}\neq0.$
The general solution is,%

\begin{align}
f(r)  &  =\chi+\frac{1}{30\widetilde{\alpha}_{3}}\sqrt[3]{\widetilde{\alpha
}_{3}^{2}\left[  5400C\ln r-40500m+300\sqrt{3\xi}\right]  }-\\
&  \frac{10r^{4}}{\sqrt[3]{\widetilde{\alpha}_{3}^{2}\left[  5400C\ln
r-40500m+300\sqrt{3\xi}\right]  }},\nonumber
\end{align}
where%

\[
\xi=\frac{1}{\widetilde{\alpha}_{3}}\left[  100r^{12}+27\widetilde{\alpha}%
_{3}\left(  2C\ln r-15M\right)  ^{2}\right]  .
\]
This solution asymptotically behaves as%

\begin{equation}
f(r)\simeq\chi.
\end{equation}
This particular case is important as far as the effect of the third order
Lovelock parameter is concerned. Hence, its physical interpretation will be
discussed in the next section.

\textit{Case 3}: \ $\widetilde{\alpha}_{2}\neq0$ and $\widetilde{\alpha}%
_{3}=0.$ The general solution is,%

\begin{equation}
f(r)=\chi+\frac{1}{10\widetilde{\alpha}_{2}r}\left(  5r^{3}\pm\sqrt
{25r^{6}-40\widetilde{\alpha}_{2}C\ln r+300\widetilde{\alpha}_{2}m}\right)  ,
\end{equation}
The asymptotic behavior is%

\begin{equation}
f(r)\simeq\chi+\frac{1\pm1}{5\widetilde{\alpha}_{2}}r^{2}.
\end{equation}

\textit{Case 4}: $3\widetilde{\alpha}_{3}=\widetilde{\alpha}_{2}^{2}.$ The
general solution for this case is,%

\begin{equation}
f(r)=\chi+\frac{r^{2}}{\widetilde{\alpha}_{2}}+\frac{1}{\widetilde{\alpha}%
_{2}}\sqrt[3]{\frac{6}{5}\widetilde{\alpha}_{2}C\ln r-r^{6}-9m\widetilde
{\alpha}_{2}},
\end{equation}
whose asymptotic behavior is,%

\begin{equation}
f(r)\simeq\chi.
\end{equation}

\subsection{The case for $k=\frac{1}{5}.$}

Another interesting case occurs for $k=\frac{1}{5}$ which is a subclass of
class \textit{A}$:$ In order to study its physical properties we first look
for the location of horizons. We have already remarked before that this
particular class satisfies only the WEC ( see Appendix) while it violates the
other energy and causality conditions. For this reason we just wish to mention
the existence of such a class without further investigation. For this
particular case, the metric function given in Eq.(22) becomes,%

\begin{equation}
f(r)=\chi+\frac{r^{2}}{\widetilde{\alpha}_{2}}\left\{  1-\sqrt[3]%
{1-\frac{3C\widetilde{\alpha}_{2}}{5r^{4}}+\frac{3m\widetilde{\alpha}_{2}%
}{5r^{6}}}\right\}  ,
\end{equation}
whose radius of horizon is obtained from Eq.(23) as%

\begin{equation}
r_{h}=\sqrt{\frac{\tilde{Q}\pm\sqrt{\tilde{Q}^{2}+20\left(  m-m_{c}\right)  }%
}{10}}%
\end{equation}
where
\begin{equation}
m_{c}=\frac{5}{3}\widetilde{\alpha}_{2}^{2},\text{ \ \ \ }\tilde{Q}=-\left(
C+5\widetilde{\alpha}_{2}\right)  .
\end{equation}

We wish to remind that, with this particular choice of $k$, the resulting
solution satisfies only the WEC. This condition further implies that $C\leq0.$
These limitations induce a number of \ possible cases for the constant
parameters $C$ and $\widetilde{\alpha}_{2}$ appearing in the Eq. (41). These
particular cases can be classified in three different classes such as
$\tilde{Q}>0,\tilde{Q}=0$ and $\tilde{Q}<0$, which we shall not consider here
any further.

\subsubsection{Naked Singularities}

In this subsection, we wish to emphasize another important property of the
Lovelock theory. We stated previously that in some special cases no horizon
forms so that the singularity at $r=0$ becomes naked with a timelike
character. In classical 4- dimensional general relativity, this is a curvature
singularity, indicating timelike geodesic incompleteness. However, our main
concern here is to analyze this naked singularity when probed with quantum
test particles. In other words, we are aiming to see whether this singular
spacetime (in classical sense) remains nonsingular quantum mechanically. To
achieve this we adopt the method initiated by Wald \cite{16} and developed by
Horowitz and Marolf \cite{17}, for static spacetimes having timelike curvature
singularities. This method states that a spacetime is quantum mechanically
nonsingular if the time evolution of any wave packet is uniquely determined by
the initial wave function. The method is briefly as follows:

A scalar quantum particle with mass M is described by the Klein-Gordon
equation $\left(  \nabla^{\mu}\nabla_{\mu}-M^{2}\right)  \psi=0.$ This
equation can be written by splitting the temporal and spatial portion as
$\frac{\partial^{2}\psi}{\partial t^{2}}=-A\psi,$ such that the spatial
operator $A$ is defined by $A=-\sqrt{f}D^{i}\left(  \sqrt{f}D_{i}\right)
+fM^{2},$ where $f=-\xi^{\mu}\xi_{\mu}$ with $\xi^{\mu}$ the timelike Killing
field, while$\ D_{i}$ is the spatial covariant derivative defined on the
static slice $\Sigma.$ \ The method requires essential self-adjointness of the
spatial operator $A$ . That is, a \textit{unique} extension of the operator
$A_{E}$ . Then, the Klein-Gordon equation for a free relativistic particle
satisfies $i\frac{\partial\psi}{\partial t}=\sqrt{A_{E}}\psi,$ with the
solution $\psi\left(  t\right)  =\exp\left(  it\sqrt{A_{E}}\right)
\psi\left(  0\right)  .$ The ambiguity occurs in the future time evolution of
the wave function ( $\psi\left(  t\right)  =\exp\left(  it\sqrt{A_{E}}\right)
\psi\left(  0\right)  $ )$,$ if $A_{E}$ is not essentially self-adjoint.
Consequently, a sufficient condition for the operator $A$ to be essentially
self-adjoint is to analyse the solutions satisfying,
\begin{equation}
A\psi\pm i\psi=0.
\end{equation}
The separable solution to Eq.(43) is assumed in the form of $\psi=\phi
(r)Y($angles$)$. The radial part becomes,%

\begin{equation}
\frac{\partial^{2}\phi}{\partial r^{2}}+\frac{1}{fr^{5}}\frac{\partial\left(
fr^{5}\right)  }{\partial r}\frac{\partial\phi}{\partial r}-\frac{c}{fr^{2}%
}\phi-\frac{M^{2}}{f}\phi\pm i\frac{\phi}{f^{2}}=0,
\end{equation}
in which $c\geq0$ is the eigenvalue of the Laplacian on the $5$-sphere.
Equation (44) can be solved with the help of Fuchsian equation \cite{17}\ by
assuming the massless case (i.e. $M=0)$ and ignoring the term $\pm i\frac
{\phi}{f^{2}}$ (since it is negligible near the origin, $r=0$). The Fuchsian
equation is $\frac{\partial^{2}\phi}{\partial r^{2}}+r^{-1}p(r)\frac
{\partial\phi}{\partial r}+r^{-2}q(r)\phi=0,$ such that $p(r)$ and $q(r)$ are
analytic at the origin. This equation admits solution in the form of
$\phi(r)=r^{\beta}\digamma(r)$, where $\digamma(r)$ is an analytic function
and $\beta$ is a complex number that solves the indicial equation $\beta
(\beta-1)+\beta p(0)+q(0)=0.$ Substituting Eq.(40) in Eq.(44), we find that
$p(0)=5$ and $q(0)=-\frac{c}{1-\left(  \frac{3m}{5\widetilde{\alpha}_{2}%
}\right)  ^{1/3}}.$ For $c=0$ (corresponds to $S$-wave)$,$ one of the two
solutions to indicial equation, solves the Eq.(44) and the resulting solution
diverges as fast as $\mid\phi\left(  r\right)  \mid^{2}=r^{-8}.$ This solution
always has infinite norm near $r=0$ since%

\begin{equation}
<\phi\mid\phi>=\int\frac{\mid\phi\left(  r\right)  \mid^{2}r^{5}}{f}dr.
\end{equation}
Consequently, $\phi\left(  r\right)  $ fails to be square integrable near the
origin. This divergence of the norm creates an infinite repulsive barrier so
that any particle remains away, and in the safer region from the origin. For
further detail in this regard we refer to \cite{18}. According to the Horowitz
- Marolf criteria, the timelike curvature singularity at the origin turns out
to be quantum mechanically nonsingular when probed with quantum test
particles. Similar analysis is also shown in Ref. \cite{3} for the
$5-$dimensional Boulware-Deser metric which also remains regular when tested
by quantum probes. A similar proof of quantum regularity applies for $N>5$ as well.

\section{Generalization to the case for higher order Lovelock theory}

In this section we give a generalization for the Lovelock gravity in higher
order. To do so, we start with an action in the form of
\begin{equation}
S=\int dx^{n+2}\sqrt{-g}\left\{  -\frac{n\left(  n+1\right)  }{3}\Lambda+%
%TCIMACRO{\tciLaplace}%
%BeginExpansion
\mathcal{L}%
%EndExpansion
_{1}+\alpha_{2}%
%TCIMACRO{\tciLaplace}%
%BeginExpansion
\mathcal{L}%
%EndExpansion
_{2}+\alpha_{3}%
%TCIMACRO{\tciLaplace}%
%BeginExpansion
\mathcal{L}%
%EndExpansion
_{3}+\alpha_{4}%
%TCIMACRO{\tciLaplace}%
%BeginExpansion
\mathcal{L}%
%EndExpansion
_{4}+...+\alpha_{\left[  \frac{n+1}{2}\right]  }%
%TCIMACRO{\tciLaplace}%
%BeginExpansion
\mathcal{L}%
%EndExpansion
_{\left[  \frac{n+1}{2}\right]  }\right\}  +S_{matter},
\end{equation}
where
\begin{equation}%
%TCIMACRO{\tciLaplace}%
%BeginExpansion
\mathcal{L}%
%EndExpansion
_{n}=2^{-n}\delta_{c_{1}d_{1}...c_{n}d_{n}}^{a_{1}b_{1}...a_{n}b_{n}%
}R_{\ \ a_{1}b_{1}}^{c_{1}d_{1}}...R_{\ \ a_{n}b_{n}}^{c_{n}d_{n}},\text{
\ \ \ }n\geq1,
\end{equation}
and the bracket $[.]$ refers to integer part. As before, Einstein equation
reads%
\begin{equation}
\mathcal{G}_{\mu}^{\nu}=\frac{n\left(  n+1\right)  }{6}\Lambda\delta_{\mu
}^{\nu}+G_{\mu}^{\nu\left(  1\right)  }+\alpha_{2}G_{\mu}^{\nu\left(
2\right)  }+\alpha_{3}G_{\mu}^{\nu\left(  3\right)  }+...+\alpha_{\left[
\frac{n+1}{2}\right]  }G_{\mu}^{\nu\left(  \left[  \frac{n+1}{2}\right]
\right)  }=T_{\mu}^{\nu}.
\end{equation}
Our static spherically symmetric metric is given by (6) which after we
rewrite
\begin{equation}
f\left(  r\right)  =\chi-r^{2}F\left(  r\right)  ,
\end{equation}
the $tt$ component of (48) becomes%
\begin{equation}
-\frac{\Lambda}{3}+F+\tilde{\alpha}_{2}F^{2}+\tilde{\alpha}_{3}F^{3}%
+...+\tilde{\alpha}_{\left[  \frac{n+1}{2}\right]  }F^{\left[  \frac{n+1}%
{2}\right]  }=\frac{M}{r^{1+n}}-\frac{2}{nr^{1+n}}%
%TCIMACRO{\tint }%
%BeginExpansion
{\textstyle\int}
%EndExpansion
r^{n}T_{t}^{t}dr,
\end{equation}
in which $M$ is an integration constant and%
\begin{equation}
\tilde{\alpha}_{s}=\overset{2s}{\underset{i=3}{\Pi}}\left(  n+2-i\right)
\alpha_{s}.
\end{equation}
$T_{t}^{t}$ is given by (15) which leads to
\begin{equation}
-\frac{\Lambda}{3}+F+\tilde{\alpha}_{2}F^{2}+\tilde{\alpha}_{3}F^{3}%
+...+\tilde{\alpha}_{\left[  \frac{n+1}{2}\right]  }F^{\left[  \frac{n+1}%
{2}\right]  }=\left\{
\begin{tabular}
[c]{ll}%
$\frac{M}{r^{1+n}}-\frac{2C}{n\left(  nk+1\right)  }r^{n\left(  k-1\right)  }$
& $nk+1\neq0$\\
$\frac{M}{r^{1+n}}-\frac{2C}{n}\frac{\ln r}{r^{1+n}}$ & $nk+1=0$%
\end{tabular}
\ \ \ \ \right.  .
\end{equation}
Here we would like to set the coefficients as
\begin{equation}
\tilde{\alpha}_{s}=\frac{\bar{\alpha}_{s}}{\bar{\alpha}_{1}},\text{ for }%
s\geq2\text{ and }-\frac{\Lambda}{3}=\frac{\bar{\alpha}_{0}}{\bar{\alpha}_{1}%
},
\end{equation}
which leads to
\begin{equation}
\sum_{s=0}^{\left[  \frac{n+1}{2}\right]  }\bar{\alpha}_{s}F^{s}=\bar{\alpha
}_{1}\times\left\{
\begin{tabular}
[c]{ll}%
$-\frac{M}{r^{1+n}}-\frac{2C}{n\left(  nk+1\right)  }r^{n\left(  k-1\right)
}$ & $nk+1\neq0$\\
$-\frac{M}{r^{1+n}}-\frac{2C}{n}\frac{\ln r}{r^{1+n}}$ & $nk+1=0$%
\end{tabular}
\ \ \ \ \right.
\end{equation}
and then we choose a specific case
\begin{equation}
\bar{\alpha}_{s}=\left(  \pm1\right)  ^{s+1}\binom{\left[  \frac{n+1}%
{2}\right]  }{s}\ell^{2s-\Delta}%
\end{equation}
where $-\frac{\Lambda}{3}=\frac{\bar{\alpha}_{0}}{\bar{\alpha}_{1}}=\pm
\frac{\ell^{-2}}{\left[  \frac{n+1}{2}\right]  }.$Following this, Eq. (52)
gives
\begin{equation}
\left(  1\pm\ell^{2}F\right)  ^{\left[  \frac{n+1}{2}\right]  }=\pm
\ell^{\Delta}\bar{\alpha}_{1}\times\left\{
\begin{tabular}
[c]{ll}%
$\frac{M}{r^{1+n}}-\frac{2C}{n\left(  nk+1\right)  }r^{n\left(  k-1\right)  }$
& $nk+1\neq0$\\
$\frac{M}{r^{1+n}}-\frac{2C}{n}\frac{\ln r}{r^{1+n}}$ & $nk+1=0$%
\end{tabular}
\ \ \ \ \right.
\end{equation}
and consequently%
\begin{equation}
f\left(  r\right)  =\chi\pm\frac{r^{2}}{\ell^{2}}\mp\frac{r^{2}}{\ell^{2}%
}\left(  \pm\left[  \frac{n+1}{2}\right]  \ell^{2}\times\left\{
\begin{tabular}
[c]{ll}%
$\frac{M}{r^{1+n}}-\frac{2C}{n\left(  nk+1\right)  }r^{n\left(  k-1\right)  }$
& $nk+1\neq0$\\
$\frac{M}{r^{1+n}}-\frac{2C}{n}\frac{\ln r}{r^{1+n}}$ & $nk+1=0$%
\end{tabular}
\ \ \ \ \right.  \right)  ^{1/\left[  \frac{n+1}{2}\right]  }.
\end{equation}
After this general solution we specify the solution for even and odd
dimensions separately. To do so, we put $\left[  \frac{n+1}{2}\right]
=\frac{n+1}{2}$ for odd dimensions and $\left[  \frac{n+1}{2}\right]
=\frac{n}{2}$ for even dimensions into (57) to obtain
\begin{equation}
f_{even}\left(  r\right)  =\chi\pm\frac{r^{2}}{\ell^{2}}\mp\left(  \pm\frac
{n}{2}\frac{1}{\ell^{n-2}}\left\{
\begin{tabular}
[c]{ll}%
$\frac{M}{r}-\frac{2C}{n\left(  nk+1\right)  }r^{nk}$ & $nk+1\neq0$\\
$\frac{M}{r}-\frac{2C}{n}\frac{\ln r}{r}$ & $nk+1=0$%
\end{tabular}
\ \ \ \ \right.  \right)  ^{2/n}%
\end{equation}
and
\begin{equation}
f_{odd}\left(  r\right)  =\chi\pm\frac{r^{2}}{\ell^{2}}\mp\left(  \pm
\frac{n+1}{2}\frac{1}{\ell^{n-1}}\left\{
\begin{tabular}
[c]{ll}%
$M-\frac{2C}{n\left(  nk+1\right)  }r^{nk+1}$ & $nk+1\neq0$\\
$M-\frac{2C}{n}\ln r$ & $nk+1=0$%
\end{tabular}
\ \ \ \ \right.  \right)  ^{2/\left(  n+1\right)  }.
\end{equation}
The latter two solutions are nothing but the BI and CS solutions \cite{13}. It
is observed that fractional powers on the paranthesis put severe restrictions
on the parameters. As a final remark in this section we note with reference to
\cite{22} that as long as our source contains an abelian gauge field such as
electromagnetism the static solution obeys the Birkhoff's theorem. For a
non-abelian gauge field, however, the problem remains open for a general
proof, which will be considered in the future separately.

\section{Generating BR Type Solutions in EGB Theory and the Theorem in
Scalar-Tensor Theory}

\subsection{BR type solutions}

Closely related with the black hole solutions is the class of BR type
solutions in GB gravity. This class arises as a limiting case of extremal
black holes so that an analogous theorem can be stated to cover this class as
well. In $N=4$, the BR solution is the unique, conformally flat EM solution.
In higher dimensions ($N>4$) we found that both, conformal and asymptotical
flatness fail\cite{15}. Being almost as important as black holes, specifically
in supergravity, we wish to present conditions on the energy momentum involved
in order to generate solutions of BR form in the EYMGB theory.

To this end we adopt the metric in the form%

\begin{equation}
ds^{2}=-fdt^{2}+f^{-1}dr^{2}+h^{2}d\Omega_{N-2}^{2},
\end{equation}
where $h$ is a constant to be specified and $f=f(r)$ is a function to be
found. The energy momentum tensor is assumed in the form%

\begin{equation}
T_{b}^{a}=C\left[  -1,-1,k,k,...\right]
\end{equation}
where $C$ and $k$ are constants that characterize the matter fields. This
energy-momentum satisfies the WEC, SEC and DEC conditions (see Appendix)
provided $C$ $\geq0$ and $k\geq\frac{1}{N-2}.$ In order to satisfy also the
causality condition we must have $C$ $\geq0$ and $\frac{1}{N-2}\leq k$ $<$
$\frac{2}{N-2}.$The Einstein's tensor in $N$-dimensions is given by,%

\begin{equation}
G_{b}^{a}=\left[  -\frac{\left(  N-3\right)  \left(  N-2\right)  }{2h^{2}%
},-\frac{\left(  N-3\right)  \left(  N-2\right)  }{2h^{2}},-\frac{\left(
N-3\right)  \left(  N-4\right)  +f^{^{\prime\prime}}h^{2}}{2h^{2}},...\right]
.
\end{equation}
in which the higher terms repeat the third one. From the Einstein's equation
$G_{ab}=T_{ab}$, we have,%

\begin{equation}
2Ch^{4}+\left(  N-3\right)  \left(  N-2\right)  h^{2}+\widetilde{\alpha
}\left(  N-5\right)  \left(  N-4\right)  =0,
\end{equation}
where $\widetilde{\alpha}=\alpha\left(  N-3\right)  \left(  N-2\right)  ,$ and%

\begin{equation}
\left(  h^{2}+2\widetilde{\alpha}^{2}\right)  f^{\prime\prime}-\left(
N-3\right)  \left(  N-4\right)  -\left(  N-5\right)  \left(  N-6\right)
\widetilde{\alpha}=2kC
\end{equation}
Solving this with the help of Eq.(63) we get,%

\begin{equation}
f(r)=\frac{\left(  2Ck+\left(  N-3\right)  \left(  N-4\right)  +\left(
N-5\right)  \left(  N-6\right)  \widetilde{\alpha}\right)  }{2\left(
h^{2}+2\widetilde{\alpha}^{2}\right)  }r^{2}+C_{1}r+C_{2},
\end{equation}
where $C_{1}$ and $C_{2}$ are integration constants. This general result
includes some of the well-known solutions for particular choices of $C$ and
$k.$ It can be anticipated that for $C_{1}\neq0\neq C_{2},f\left(  r\right)
=0,$ admits roots resulting in non-asymptotically flat black hole solutions.
Beside this, for $C=q^{2}$ ( $q$ is the electric charge) and $k=1$ it
corresponds to $N$ - dimensional BR like solution in the Einstein-Maxwell
theory. Note that we choose the integration constants, $C_{1}=C_{2}=0$ for
this particular case. Another interesting solution is obtained for the EYM
theory if one takes $C=\frac{\left(  N-3\right)  \left(  N-2\right)  }{2Q^{2}%
}$ and $k=-\frac{N-6}{N-2}.$

\subsection{Salgado's theorem in higher dimensions and Scalar Tensor Theory
(STT) of gravity}

We start with an action in $n+2-$dimensions\cite{21}
\begin{equation}
I=%
%TCIMACRO{\tint }%
%BeginExpansion
{\textstyle\int}
%EndExpansion
d^{n+2}x\sqrt{-g}\left\{  \frac{1}{2}F\left(  \phi\right)  R-\frac{1}%
{2}\left(  \nabla\phi\right)  ^{2}\right\}
\end{equation}
in which $\phi$ is a massless scalar field a function of only radial
coordinate $r$ and $F\left(  \phi\right)  $ is a function of $\phi$ to be
identified later. The field equations by using the usual variation method are
given by%
\begin{align}
G_{\mu\nu}  &  =T_{\mu\nu},\\
T_{\mu\nu}  &  =\frac{1}{F\left(  \phi\right)  }\left[  \nabla_{\mu}%
\nabla_{\nu}F\left(  \phi\right)  +\nabla_{\mu}\phi\nabla_{\nu}\phi-g_{\mu\nu
}\nabla^{2}F\left(  \phi\right)  -\frac{1}{2}g_{\mu\nu}\left(  \nabla
\phi\right)  ^{2}\right]  ,\\
\nabla^{2}\phi &  =-\frac{1}{2}F^{\prime}\left(  \phi\right)  R.
\end{align}
Now the trace of (68) manifests%
\begin{equation}
R=\frac{\left(  \frac{n}{2}+\left(  n+1\right)  F^{\prime\prime}\right)
\left(  \nabla\phi\right)  ^{2}}{\frac{n}{2}F+\frac{\left(  n+1\right)  }%
{2}\left(  F^{\prime}\right)  ^{2}},
\end{equation}
which, while $\nabla\phi\neq0,$ we wish to make it zero i.e.,%
\begin{equation}
\frac{n}{2}+\left(  n+1\right)  F^{\prime\prime}=0,
\end{equation}
which yields%
\begin{equation}
F\left(  \phi\right)  =-\frac{n}{4\left(  n+1\right)  }\phi^{2}+C_{1}%
\phi+C_{2}%
\end{equation}
where $C_{1}$ and $C_{2}$ are integration constants. In order to follow our
goal we set $C_{1}=0$ and $C_{2}=1$ such that%
\begin{equation}
F\left(  \phi\right)  =1-\frac{n}{4\left(  n+1\right)  }\phi^{2}.
\end{equation}
We put these results into the field equations%
\begin{equation}
\nabla^{2}\phi=0
\end{equation}
which gives%
\begin{equation}
\partial_{r}\left(  r^{n}\frac{N}{A}\phi_{r}\right)  =0
\end{equation}
and%
\begin{equation}
T_{\mu\nu}=\frac{1}{F\left(  \phi\right)  }\left[  \left(  1-2\zeta\right)
\nabla_{\mu}\phi\nabla_{\nu}\phi+\left(  2\zeta-\frac{1}{2}\right)  g_{\mu\nu
}\nabla^{\alpha}\phi\nabla_{\alpha}\phi-2\zeta\phi\nabla_{\mu}\nabla_{\nu}%
\phi\right]
\end{equation}
where $\zeta=\frac{n}{4\left(  n+1\right)  }.$ The later expression directly
gives
\begin{equation}
T_{r}^{r}=\frac{1}{FA^{2}}\left[  \frac{1}{2}\phi_{r}^{2}+2\zeta\phi\phi
_{r}\frac{A_{r}}{A}-2\zeta\phi\phi_{rr}\right]  ,
\end{equation}
and
\begin{equation}
T_{t}^{t}=-\frac{\phi_{r}}{FA^{2}}\left[  \left(  \frac{1}{2}-2\zeta\right)
\phi_{r}+2\zeta\phi\frac{N_{r}}{N}\right]  .
\end{equation}
Now we use (75) to get $\phi_{rr}=-\phi_{r}\left[  \partial_{r}\ln\left(
r^{n}\frac{N}{A}\right)  \right]  $ and then%
\begin{equation}
T_{r}^{r}-T_{t}^{t}=\frac{n+2}{2\left(  n+1\right)  }\frac{\phi_{r}}{FA^{2}%
}\frac{1}{\left(  r^{n}N^{2}\right)  ^{\frac{n}{n+2}}}\partial_{r}\left[
\phi\left(  r^{n}N^{2}\right)  ^{\frac{n}{n+2}}\right]  ,
\end{equation}
which, as the first requirement in our theorem, must be zero i.e.,%
\begin{equation}
\partial_{r}\left[  \phi\left(  r^{n}N^{2}\right)  ^{\frac{n}{n+2}}\right]
=0,
\end{equation}
or%
\begin{equation}
\phi\left(  r^{n}N^{2}\right)  ^{\frac{n}{n+2}}=d,
\end{equation}
where $d$ is a constant. Also after knowing $A=\frac{1}{N}$ (this was proved
before), from (75), one gets%
\begin{equation}
\phi_{r}=a\frac{A^{2}}{r^{n}},
\end{equation}
which together with (81) admit a solution for scalar field as
\begin{align}
\phi &  =\left(  -\frac{e_{n}}{r-M}\right)  ^{\frac{n}{2}},\\
N^{2}  &  =\frac{1}{A^{2}}=\frac{d^{1+\frac{2}{n}}}{\left(  -e_{n}\right)
^{1+\frac{n}{2}}}r^{1-\frac{n}{2}}\left(  1-\frac{M}{r}\right)  ^{1+\frac
{n}{2}},
\end{align}
where $e_{n}=\frac{n}{2}\frac{d^{1+\frac{2}{n}}}{a}$ and $M$ is a constant.
These result help us to find the closed form of $T_{r}^{r}=T_{t}^{t}$ and
$T_{\theta}^{\theta}$ as%
\begin{align}
T_{r}^{r}  &  =T_{t}^{t}=\frac{\phi_{r}}{2\left(  n+1\right)  FA^{2}}\frac
{n}{2}\left[  \phi_{r}+\frac{n}{r}\phi\right]  ,\\
T_{\theta}^{\theta}  &  =-\frac{\phi_{r}}{2\left(  n+1\right)  FA^{2}}\left[
\phi_{r}+\frac{n}{r}\phi\right]
\end{align}
which after considering $k=\frac{T_{\theta}^{\theta}}{T_{t}^{t}}$ (from the
theorem) one finds%
\begin{equation}
k=\frac{T_{\theta}^{\theta}}{T_{t}^{t}}=-\frac{2}{n}.
\end{equation}
The last requirement to fulfill the conditions in the theorem, is to adjust
the free parameters such that $T_{r}^{r}=T_{t}^{t}\sim\frac{1}{r^{n\left(
1-k\right)  }}=\frac{1}{r^{n+2}}$ i.e.,
\begin{equation}
\frac{\phi_{r}}{2\left(  n+1\right)  FA^{2}}\frac{n}{2}\left[  \phi_{r}%
+\frac{n}{r}\phi\right]  =\frac{C}{r^{n+2}}%
\end{equation}
which after substitution the closed form of all functions on gets%
\begin{equation}
\frac{\phi\left[  \frac{1}{2e_{n}}\phi^{\frac{2}{n}}+\frac{1}{r}\right]
}{1-\frac{n}{4\left(  n+1\right)  }\phi^{2}}=\frac{4C\left(  n+1\right)
}{an^{2}r^{2}}%
\end{equation}
or simply%
\begin{equation}
\frac{r\left(  r-2M\right)  }{\left[  \left(  r-M\right)  ^{1+\frac{n}{2}%
}-\frac{n}{4\left(  n+1\right)  }\left(  -e_{n}\right)  ^{n}\left(
r-M\right)  ^{1-\frac{n}{2}}\right]  }=\frac{8C\left(  n+1\right)  }%
{an^{2}\left(  -e_{n}\right)  ^{\frac{n}{2}}}=cons..
\end{equation}
As one may notice, this is a very strong condition and only in $4-$dimensions
can be satisfied, i.e., for $n=2$ we get%
\begin{equation}
\frac{r\left(  r-2M\right)  }{\left[  \left(  r-M\right)  ^{2}-\frac{1}%
{6}e^{2}\right]  }=-\frac{6C}{d^{2}}%
\end{equation}
which gives%
\begin{equation}
d^{2}=-6C\text{ \ \ and \ \ }e^{2}=6M^{2}.
\end{equation}
Nevertheless one finds%
\begin{equation}
T_{r}^{r}=T_{t}^{t}=-\frac{d^{2}}{6r^{4}}%
\end{equation}
and%
\begin{align}
\phi &  =\left(  -\frac{e}{r-M}\right)  ,\\
N(r)^{2}  &  =A(r)^{-2}=\chi-\frac{2m}{r}-\frac{C}{r^{2}}=\frac{d^{2}}{e^{2}%
}\left(  1-\frac{M}{r}\right)  ^{2}.
\end{align}
For $S_{2}$ i.e., $\chi=1$ one gets $d^{2}=e^{2}=6M^{2},$ $m=M$ and $C=-M^{2}$
which reveal%
\begin{equation}
T_{r}^{r}=T_{t}^{t}=-\frac{M^{2}}{r^{4}},\text{ \ \ }N(r)^{2}=A(r)^{-2}%
=\left(  1-\frac{M}{r}\right)  ^{2},
\end{equation}%
\begin{equation}
\phi=\left(  \pm\frac{\sqrt{6}M}{r-M}\right)  .
\end{equation}
For $H_{2}$ ($\chi=-1$) we find $\frac{d^{2}}{e^{2}}=-1$, $m=-M$ and $C=M^{2}$
which means%
\begin{align}
T_{r}^{r}  &  =T_{t}^{t}=\frac{M^{2}}{r^{4}},\text{ \ \ }N(r)^{2}%
=A(r)^{-2}=-\left(  1-\frac{M}{r}\right)  ^{2},\\
\phi &  =\left(  \pm\frac{\sqrt{6}M}{r-M}\right)  .
\end{align}
Finally for $\chi=0$ this solution is not applicable.

\section{Conclusion}

In this paper, we have extended the Salgado's theorem to generate static,
spherically symmetric black hole solutions in higher dimensional Lovelock
gravity with matter fields. We have shown explicitly that our general solution
recovers formerly obtained solutions in particular limits. A new class of
black hole solutions in $7$-dimensions known as Chern-Simon black holes with
specific matter fields is presented in detail. The matter fields couple to the
system through the constant parameters $C$ and $k$. It is shown that these
parameters are restricted as a result of energy conditions.

Before attempting the most general solution, firstly, we derive the general
form of the $N$- dimensional black hole solutions in the third order Lovelock
gravity. Due to the technical reasons, we constraint, the parameters
$\widetilde{\alpha}_{2}$ and $\widetilde{\alpha}_{3}$ as in Ref. \cite{4}, so
that we obtain solutions that overlap with the known solutions in $7$- dimensions.

A new black hole solution in $7$-dimensions with $k=\frac{1}{5}$ is obtained.
It is shown that depending upon the values of $C$ and $\widetilde{\alpha}_{2}%
$, one or two horizons may develop. Asymptotically, depending on the
parameters, our new solutions are either flat or de Sitter/anti- de Sitter
types. Under some special conditions the \textit{naked} singularity becomes
inevitable. In classical $4$- dimensional general relativity, this singularity
is a timelike curvature singularity. The structure of this singularity is
further analyzed by quantum test particles according to the method developed
by Horowitz and Marolf \cite{17}. Our analysis has revealed that, although
$r=0$ is singular in classical sense, it becomes nonsingular when probed with
quantum test particles. It has also been shown for $N=7$ that the third order
Lovelock parameter plays an effective role in removing the black hole property
and leaving the singularity at $r=0$ as naked. Higher Lovelock parameters,
$\alpha_{s},$ $\left(  s>3\right)  $ play the similar role for $N>7$. Under
the light of these results, the Lovelock theory of gravity becomes important
in the sense that, it provides an arena to investigate the contribution of
higher order curvature terms at short distances; especially for the solutions
that incorporate black holes. Our final remark is to extend the theorem to
cover Bertotti-Robinson type solutions and scalar-tensor theory. It is found
that in the scalar-tensor theory Salgado's theorem for $N>4$ does not work.

\bigskip

\textbf{APPENDIX A: Energy Conditions}

When a matter field couples to any system, energy conditions must be satisfied
for physically acceptable solutions. We follow the steps as given in \cite{8}.

\subsection{Weak Energy Condition (WEC)}

The WEC states that,%

\begin{equation}
\rho\geq0\text{ \ \ \ \ \ \ \ \ \ \ \ and \ \ \ \ \ \ \ \ }\rho+p_{i}%
\geq0\text{ \ \ \ \ \ ( }i=1,2,...n+1) \tag{A1}%
\end{equation}
in which $\rho$ is the energy density and $p_{i}$ are the principal pressures
given by%

\begin{equation}
\rho=-T_{t}^{t}=-T_{r}^{r}=-\frac{C}{r^{n(1-k)}},\text{ \ \ \ \ \ \ \ \ }%
p_{i}=T_{i}^{i}\text{ \ \ (no sum convention)} \tag{A2}%
\end{equation}
The WEC imposes the following conditions on the constant parameters $C$ and
$k$;%

\begin{equation}
C\leq0\text{ \ \ \ \ \ and \ \ \ \ \ }k\leq1, \tag{A3}%
\end{equation}

\subsection{Strong Energy Condition (SEC)}

This condition states that;%

\begin{equation}
\rho+%
%TCIMACRO{\dsum \limits_{i=1}^{n+1}}%
%BeginExpansion
{\displaystyle\sum\limits_{i=1}^{n+1}}
%EndExpansion
p_{i}\geq0\text{ \ \ \ \ \ \ and \ \ \ \ \ \ }\rho+p_{i}\geq0. \tag{A4}%
\end{equation}
This condition together with the WEC constrain the parameters as,%

\begin{equation}
C\leq0\text{ \ \ \ \ \ and \ \ \ \ \ }k\leq0. \tag{A5}%
\end{equation}

\subsection{Dominant Energy Condition (DEC)}

In accordance with DEC, the effective pressure $p_{eff}$ should not be
negative i.e. $p_{eff}\geq0$ where%

\begin{equation}
p_{eff}=\frac{1}{n+1}%
%TCIMACRO{\dsum \limits_{i=1}^{n+1}}%
%BeginExpansion
{\displaystyle\sum\limits_{i=1}^{n+1}}
%EndExpansion
T_{i}^{i}=-\frac{\left(  1+nk\right)  }{1+n}\rho. \tag{A6}%
\end{equation}
One can show that DEC, together with SEC and WEC impose the following
conditions on the parameters
\begin{equation}
C\leq0\text{ \ \ \ \ \ and \ \ \ \ \ }-1\leq k\leq0. \tag{A7}%
\end{equation}
It is observed that the simplest case is provided by $k=-\frac{1}{n}$ (class
B) which yields $p_{eff}=0.$

\subsection{Causality Condition}

In addition to the energy conditions one can impose the causality condition%

\begin{equation}
0\leq\frac{p_{eff}}{\rho}<1, \tag{A8}%
\end{equation}
which implies%

\begin{equation}
C\leq0\text{ \ \ \ \ \ and \ \ \ \ \ }-1\leq k\leq-\frac{1}{n} \tag{A9}%
\end{equation}
Our set of class B solutions automatically satisfy the causality condition.
Concerning the class A solutions some members, such as $k=\frac{1}{n}$
violates causality.

\begin{center}

\end{center}

\end{document}